# Vladimir Platonovich Tsesevich (11.10.1907 – 28.10.1983)

Ivan L. Andronov

Department "Mathematics, Physics and Astronomy", Odessa National Maritime University

Abstract. The history of astronomy in Odessa is briefly discussed with a special emphasis on the scientific school of variable star researches founded by Prof. V.P.Tsesevich.

Biografia Vladimira Platonovicha Tsesevicha, lidera astronomii w Odessie w latach 1944-1983, została krótko omówiona, a także kierunków studiów, głównie w świetle badań gwiazd zmiennych przez członków szkoły naukowej6 założonej przez niego. Kierunki tych badań obejmują bardzo szeroki zakres typów zmienności - "magnetycznych" i "niemagnetycznych" zmiennych kataklizmicznych - symbiotycznych, rentgenowskich i innych współdziałających układów binarnych, klasycznych zaćmieniowych i "ekstremalnych bezpośrednich impaktorów", zmienne pulsujące od DSct i RR przez C i RV do SR i M. Poprawione algorytmy i programy zostały opracowane dla statystycznie optymalnego modelowania fenomenologicznego i fizycznego.

Początkowo studia te w Odessie inspirowane były Profesorem Władimirem Płatonoviczem Tsesewiczem, który był ("z wielką literą") Skrupulatnym Naukowcem i Błyskotliwym Pedagogiem, Gruntownym Autorem i Błyskotliwym Wychowawcem: inteligentnym Popularyzatorem, Wytrwałym Organizatorem i Radosnym Jokerem - prawdziwym Profesorem i Nauczycielem. Był "Poetą Gwiezdnych Niebios".

Vladimir Platonovich Tsesevich urodził się w Kijowie 11.10.1907. Jego ojciec Tsesevich Platon Ivanovich był znanym śpiewakiem operowym (głos basowy). Jego piękne piosenki można znaleźć w Internecie. Jego matka Kuznetsova Elisaveta Aleksandrovna była aktorką operową, później pedagogiem. V.P. Tsesevich rozpoczął edukację na Uniwersytecie Państwowym w Leningradzie w wieku prawie 15 lat i wybrał gwiazdy zmienne jako główny kierunek studiów.

Pierwsza praca oparta na jego spostrzeżeniach (ale nie był współautorem) pojawiła się również w 1922 roku, ale pierwsza jego praca ukazała się w przyszłym roku (całkowicie 42 podczas jego studencyjnego stażu). Nazwa została napisana jako "W.Zessewitsch" w artykułach opublikowanych w "Astronomische Nachrichten" w Niemcach.

Opiekunem jego studiów doktoranckich był G.A.Tichow, znany astronom, który studiował głównie planetę Mars. Po Leningradzie, w latach 1933-1937, był dyrektorem Tadżykistańskiego Obserwatorium Astronomicznego (obecnie Instytut Astrofizyki). W latach 1937-1942 pracował w Leningradzie. Ewakuowany do Duszanbe, gdzie pracował jako profesor, a następnie przeniósł się do Odessiego Uniwersitety Panstwowego w 1944 roku.

Pod jego nadzorem początkowo małe obserwatorium astronomiczne w Odessie stało się jedną z wiodących organizacji astronomicznych. W międzynarodowym roku geofizycznym 1957 roku powstały podmiejskie stacje astronomiczne w Majakach i Kryżanovce, później - teleskopy na Pikie Terskol (obecnie Rosja) i zadbane Duszanbe (obecnie Tadżykistan). Po śmierci V.P.Tsesevicha, 1-m lustro wykonane w Odessie, został zainstalowany w Vihorlat National Telescope (VNT) na Słowacji. Jest wykorzystywany w wielu programach międzynarodowych, także z Ukrainą i Polską. Zorganizował dziennik Izvestiya Astronomical Observatory (od 1947 r.), który od 1992 r. Jest publikowany w języku angielskim pod tytułem "Odessa Astronomical Publications" (OAP). W 2017 r. publikowany jest jubileuszowy 30 tom (http://oap.onu.edu.ua ).

W latach 1948-50, jego talent organizatora został skutecznie zrealizowany jako dyrektora Głównego Obserwatorium Astronomicznego Ukraińskiej Akademii Nauk (1948-1950). W 1948 r. został wybrany na członka korespondenta Ukraińskiej Akademii Nauk.

W naszej pamięci pozostał głębokim wrażeniem jego niezwykłej energii i entuzjazmu Władimira Płatonowicza Tsesewicza. Był nie tylko wybitnym naukowcem i organizatorem nauki, założycielem szkoły naukowej badaczy gwiazd zmiennych, ale także wybitnym specjalistą i popularyzatorem astronomii. Lista jego publikacji zawierała 759 tytułów (Dziubina & Rikun, 1988), pomiędzy nimi 22 monografie. Jednak w ADS jest tylko niewielka ich część (142 dla "Tsesevich", 31 "Tsessevich", 44 "Zessewitsch").

Jego wykłady dla studentów były urzekające i ekscytujące. Wśród nich były "Dodatkowe rozdziały matematyki matematycznej" i dwa semestry "Astrofizyki Relatywistycznej". Czytał wykłady jako wiersz i było to ekscytujące. Wydedukował wiele formuł bez notatek z wykładów, zaniedbując dzwonki, i w konsekwencji zatrzymał się, gdy trudne przejście do jakiejś formuły, czasem trwające kilka stron, dobiegło końca. Uczył nie bać się wyzwań i wykazywał proces twórczy. Dokładając się do siebie, wymagał od innych. Poszukiwał nowych odkryć i pomysłów w astronomii oraz zaproponował nowe kierunki studiów. Jego ostatnią naukową pasją był magnetyczny, kataklizmiczny układ podwójny AM Herculesa. Pierwsze artykuły ukazujące jego ekstremalny egzotyczny charakter zostały opublikowane prawie 40 lat temu (Andronov, Vasilieva & Tsesevich, 1980).

V.P.Tsesevich nadzorował ponad 40 doktorantów, tworząc skuteczną szkołę naukową i wspierał różne naukowe kierunki. V.P. Tsesevich był zupełnie inny, podobny do gwiazd zmiennych, które studiował.

Poza tym, że był dyrektorem obserwatorium astronomicznego Odesskiego Panstwowego (obecnie Narodowy) Uniwersytetu (ONU) i katedry Astronomii Wydziału Fizycznego ONU, wykładał w innych instytutach – w Institute Lodowego przemysłu, w Odesskiej Wyższej szkoły inżynierii morskiej, w Odesskim institucie inżynierów flota morskiego (obecnie Odesski Narodowy Uniwersytet Morski (ONMU)). W ostatnim instytucie wykładał przez 15 lat (1952-1967), a nawet był szefem katedry Matematyki Wyższej (obecnie katedra Matematyki, Fizyki i Astronomii).

Wśród tych publikacji znalazły się liczne monografie, które można podzielić na "drukowane katalogi" i "klasyczne" monografie, takie jak "Zmienne gwiazdy i metody ich obserwacji" (Tsesevich, 1970, 1980), kolektywne monografie " Zaćmieniowe zmienne gwiazdy"(Tsesevich, 1971),"Gwiazdy typu RR Lyrae" (Tsesevich, 1969), oba przetłumaczone na język angielski.

Najbardziej popularną książką jest "Co i jak obserwować na niebie" (Tsesevich, 1984). Został wydany sześć razy, ostatni raz w 1984 roku i jest nawet teraz wśród najlepszych książek.

Nie wszyscy czytelnicy zostali astronomami, ale wiele osób pamięta tę książkę, a wielu studentów pamięta utalentowane i emocjonalne wykłady zawodowe i publiczne. To zainicjowało pasję do astronomii (i nauki generalnie) wielu naukowców, którzy teraz stanowią podstawę naszej wspaniałej nauki.

Prof. Tsesewicz miał dobrą współpracę z kolegami z AAVSO (USA), odbył długą podróż, gdzie dokonał wielu obserwacji wizualnichh na płytach fotograficznych z największej kolekcji przechowywanej w Uniwersitecie Harvardskim.

Bardzo efektywna współpraca dotyczyła Obserwatorium Astronomiczne Uniwersytetu Jagiellonskiego w Krakowie. W 1972 roku wybitny polski astronom Prof. Kazimierz Kordylewski odwiedził Odessę, co zainicjowało wspólny kierunek badań. Specjalna grupa obserwatorów na stacji Majaki pracowała nad określeniem ekstremów grupy gwiazd typu RR Lyr z obserwacji wizualnych, fotograficznych i fotoelektrycznych, które zostały opublikowane w "Supplement" "Rocznik Astronomiczny Obserwatorium Krakowskiego" (SAC = "Supplemento Internationale de Anuario Cracoviense"). Prof. P.Flin (2007) przedstawia więcej szczegółów, np. że prof. Tsesevich (pisany po polsku jako Cesewicz) został po raz pierwszy wymieniony w SAC w 1925 roku.

Ciekawy zbieg okoliczności – obaj profesorowie urodzili się 11 października (1904 – K. Kordylewski, 1907 – V.P. Tsesevich).

Potem efektywne kontakty naukowe między astronomami z Odessy a polskimi astronomami uzyskały dalszy rozwój w kierunku badań gwiazd zmiennych (Dr. B.Wszołek, Prof. S.Zoła, Dr. W.Godłowski, Dr. R.Sczerba i in.)), a także w astronomii i astronomii - pozagalaktycznej (Prof. P.Flin) i satelitarnej (Prof.

E.Wnuk) oraz w edukacji astronomicznej (J.Jagła, PTMA; D.Pasternak w MOA - Młodzieżowe Obserwatorium Astronomiczne im. Kazimierza Kordylewskiego w Niepołomicach).

Najważniejsze elementy międzynarodowej kampanii "Inter-Longitude Astronomy" (ILA), w szczególności uzyskane w ramach tej współpracy, zostały przedstawione przez Andronova i in. (2003, 2010, 2014, 2017) i zaplanowana jest długa lista artykułów.

W 2017r. odbyła się serja wydarzeń poświęconych 100. rocznicy V.P.Tsesewicza: międzynarodowa konferencja Gamowska (ONU, Odessa, Ukraina, 13-18.08.2017, http://gamow.odessa.ua); jednodniowa konferencja "Modelowanie matematyczne, procesy fizyczne i obiekty astronomiczne" ze współautorami międzynarodowymi (ONMU, Odessa, 11.10.2017), Spotkanie Ukraińskiego Stowarzyszenia Astronomicznego (MAO, Kijów, 25.10.2017) oraz spotkanie w Odesskim Domie Naukowców (Odessa, 27.10.2017).

Wyszukiwarka Google dla V.P.Tsesevich («В.П.Цесевич» po rosyjsku) daje 33700 wpisów, z których chcielibyśmy wskazać kilka wspomnień które napisali: Volyanska, Karetnikov & Mandel (2007), Vavilova (2017), Andronov (2003, 2017), Samus ' (1988, 2007), Bezdenezhnyi (2007). Jego prawie ostatni wywiad został przygotowany przez T.M.Nepomnjashchaja (Tsesevich, 1978). Kompilacja wspomnień została przygotowana na jego 100-lecie (Tsesevich, 2007).

Na diagramie "Astronomer H-R diagram" (https://www.strudel.org.uk/blog/astro/000943.shtml), (wyniki wyszukiwania w Google od liczby artykułów naukowych), Prof. V.P.Tsesevich znajduje się w lewej górnej części, tuż pod tytułem "Akademiccy giganci".

Asteroida 2498 (odkryta 23.08.1977 przez Prof. N.S.Chernykh w Krymskim Obserwatorium Astrofizycznym) została nazwana "Tsesevich". Jak napisano na oficjalnym certyfikacie Międzynarodowej Unii Astronomicznej, "Nazwany na cześć Vladimira Platonovicha Tsesevicha (1907-1983), byłego dyrektora Obserwatorium Uniwersytetu w Odessie, znanego z badań gwiazd zmiennych. Studiował także zmiany jasności asteroid {433} Erosa i jest autorem podręcznika dla astronomów-amatorów ".

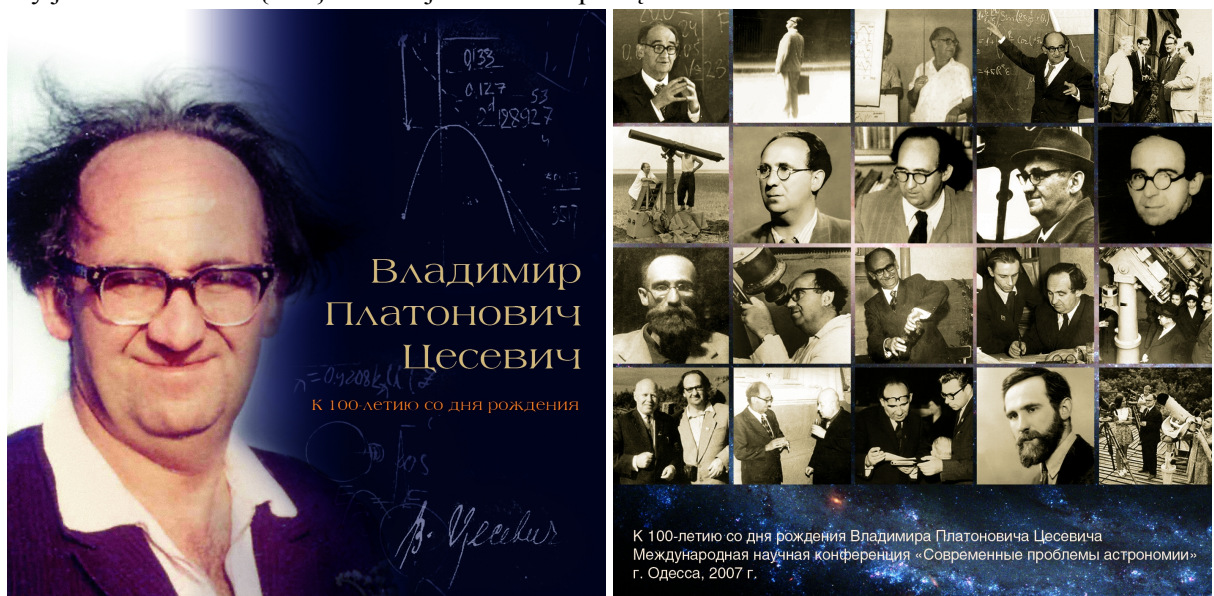

Kolaż na setną rocznicę (by Prof. V.V.Kovtyukh).

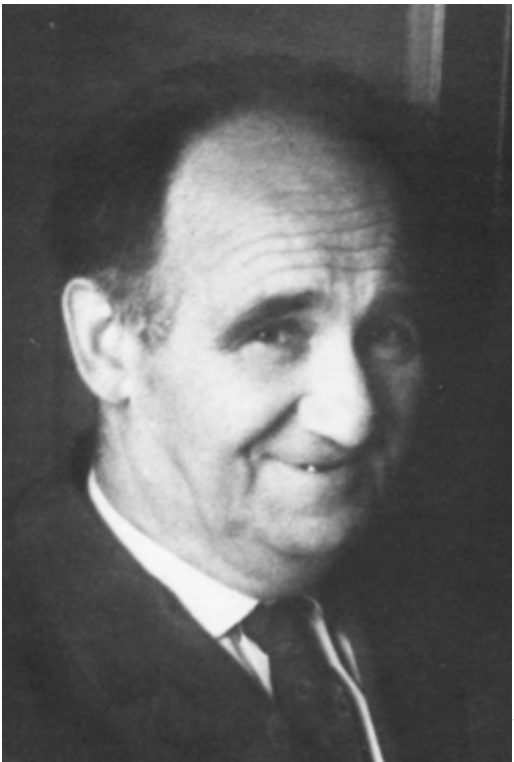
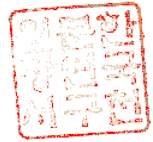
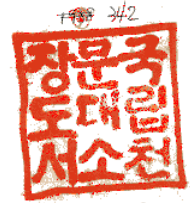

Nieformalne zdjęcie i skan najsłynniejszej książki.

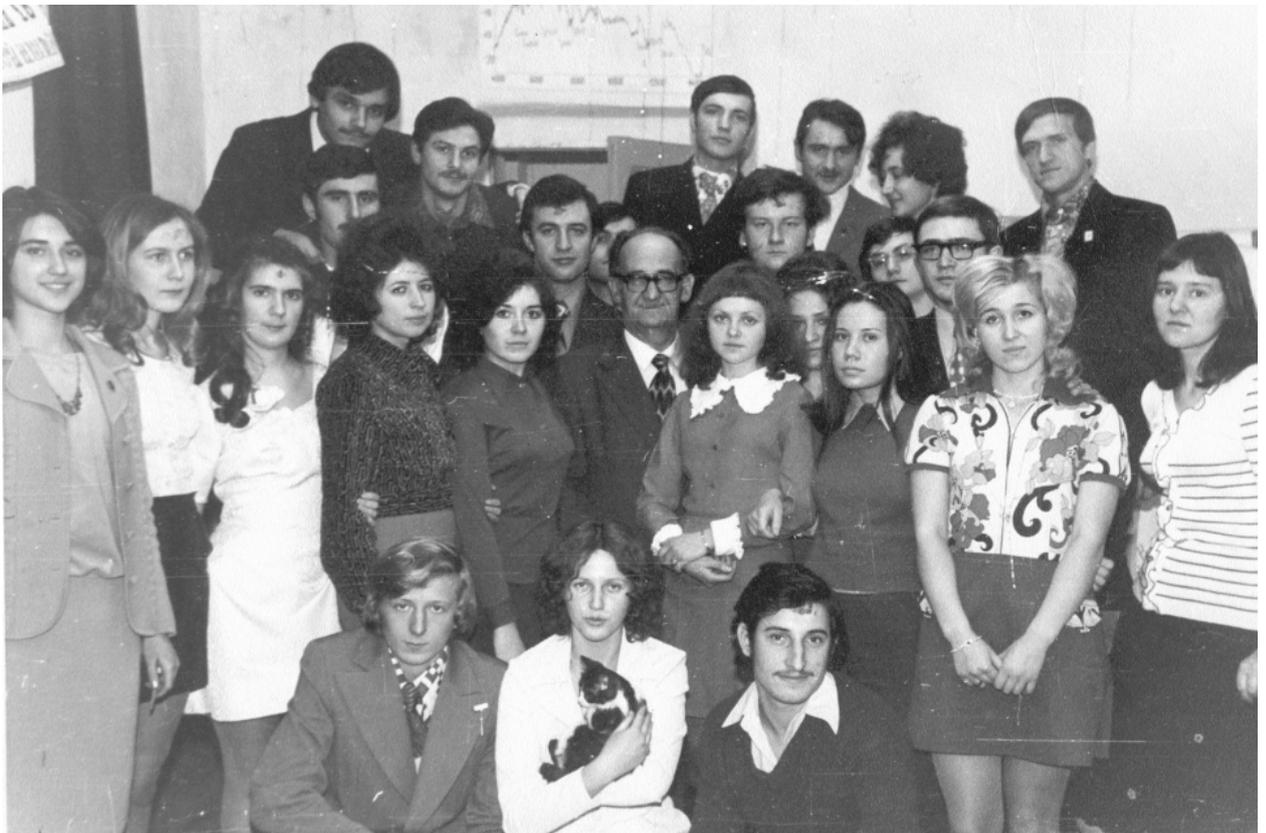

Pierwsze znaczki "prawdziwych astronomów" (1975). To była spektakl teatralny i "szybki egzamin", po którym studenci pierwszego roku "dostali astronomiczne znaczki" i "stali się prawdziwymi astronomami". Profesor jest w centrum, a autor tego artykułu jest nieco w górze i w prawo. 3 studentów na tym zdjęciu zostało Doktorami nauk fizycznych i matematycznych, a 5 innych – doktorami PhD.

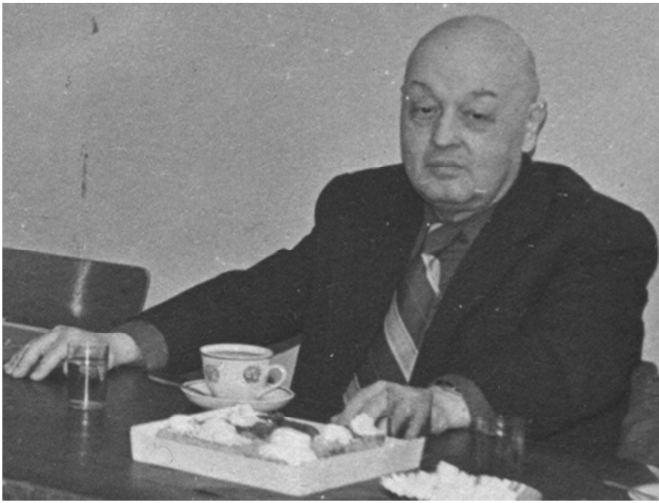
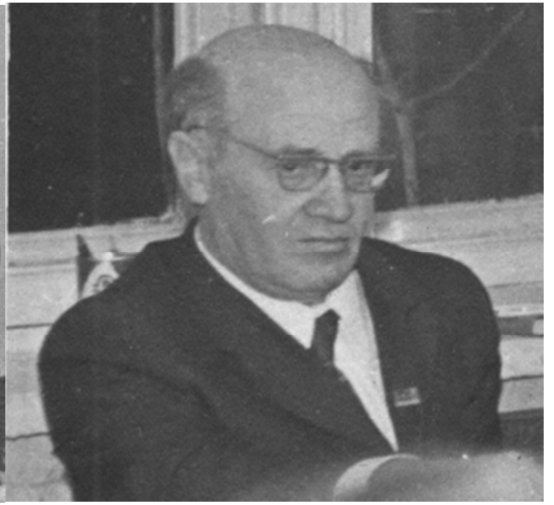
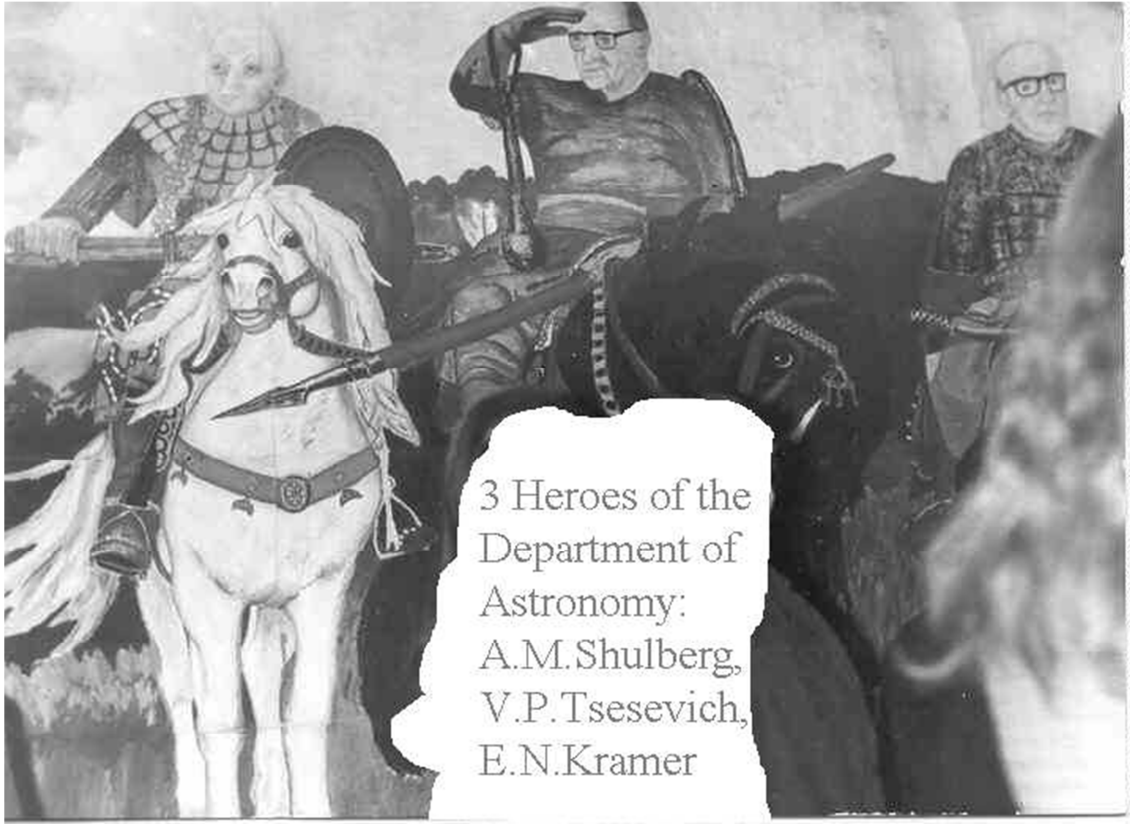

Malarstwo przez studentów. Trzy wspaniałe bohatery - czołowe nauczycieli katedry astronomii, patriarchowie.

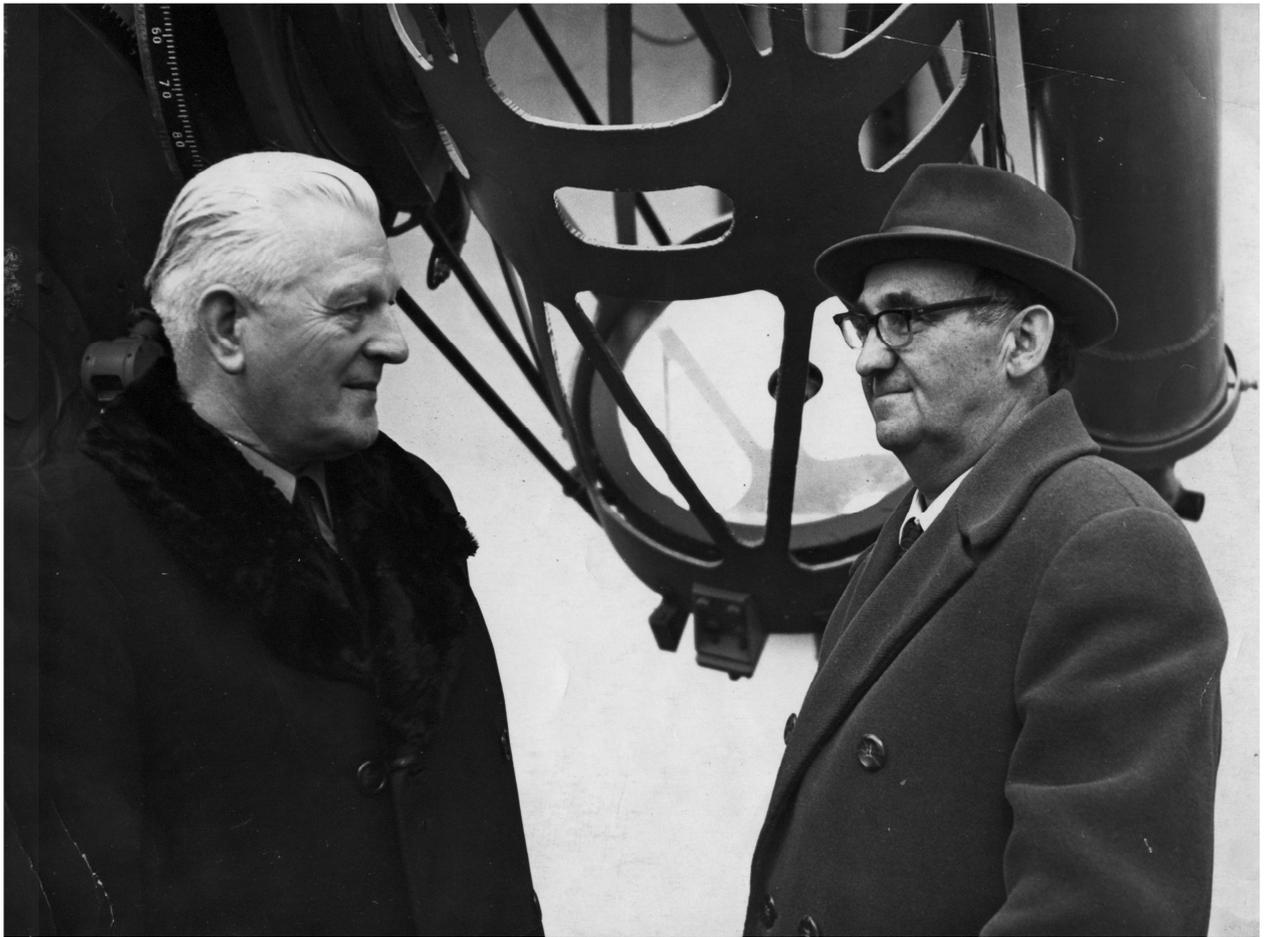

Profesor Kazimierz Kordylewski (11.10.1903-11.03.1981) i Profesor Wladimir P. Cesewicz (11.10.1907-28.10.1983), Majaki, Obserwatorium Astronomiczne Odesskiego Uniwersytetu Panstwowego (1972).